\documentclass[11pt]{iopart}

\usepackage[utf8]{inputenc} 
\usepackage[T1]{fontenc}    
\usepackage{hyperref}       
\usepackage{url}            
\usepackage{booktabs}       
\usepackage{amsfonts}       
\usepackage{nicefrac}       
\usepackage{microtype}      
\usepackage{lipsum}
\usepackage{graphicx}
\usepackage{color,soul}
\begin{document}

\title[Forecasting the outcome of COVID-19 outbreaks from fatality time series]{Forecasting the outcome and estimating the epidemic model parameters from the fatality time series in COVID-19 outbreaks  }

\author{Gábor Vattay}

\address{Department of Physics of Complex Systems, Eötvös Loránd University \\ H-1053 Budapest, Egyetem tér 1-3., Hungary}
\ead{vattay@elte.hu}
\vspace{10pt}
\begin{indented}
\item[]April 2020
\end{indented}

\begin{abstract}
In the absence of other tools, monitoring the effects of protective measures, including social distancing and forecasting the outcome of outbreaks is of immense interest. Real-time data is noisy and very often hampered by systematic errors in reporting. Detailed epidemic models may contain a large number of empirical parameters, which cannot be determined with sufficient accuracy.  
In this paper, we show that the cumulative number of deaths can be regarded as a master variable, and the parameters of the epidemic such as the basic reproduction number, the
size of the susceptible population, and the infection rate can be determined.
In the SIR model, we derive an explicit single variable differential equation for the evolution of the cumulative number of fatalities.  We show that the epidemic in Spain, Italy, and Hubei Province, China follows this
master equation closely. We discuss the relationship with the logistic growth model, and we
show that it is a good approximation when the basic reproduction number is less than $2.3$.
This condition is valid for the outbreak in Hubei, but not for the outbreaks in Spain, Italy, and New York. The difference is in the shorter infectious period in China, probably due to the separation policy of the infected. 
For more complex models, with more internal variables, such as the SEIR model, the equations derived from the SIR model remain valid approximately, due to the separation of timescales.

\end{abstract}

%
%
%
%
%

\section{Introduction}

In the absence of other tools, monitoring the effects of protective measures, including social distancing, and forecasting the outcome of outbreaks is of immense interest\cite{thelancet,vattay}. Real-time data is noisy and very often hampered by systematic errors in reporting. Detailed epidemic models may contain a large number of empirical parameters\cite{lancet}, which cannot be determined with sufficient accuracy. In this situation, low dimensional, effective models of epidemic dynamics with the least number of variables and parameters are needed for robust modeling and forecasting\cite{dyn}. Such models naturally arise in
statistical mechanics, where the behavior of coupled micro-systems with many parameters and variables, can be reduced to the dynamics of a few macroscopic variables on large scales.

In the current situation, disease control and prevention agencies publish macroscopic data such as the number of cases, recovered, and deceased patients daily with a spatial resolution of administrative districts containing millions of inhabitants. Ideally, the model should be able to work on this level of resolution. Parameters of the disease
the spreading process is also hard to determine, and even the few essential parameters have a wide range in the available literature. 

In the absence of details, compartmental epidemic models describing the average behavior of the system can be a starting point. Even the simplest
models contain several variables, which are hard to determine from
the available data. The minimal SIR model\cite{sir,Brauer} describes the behavior
of the susceptible $S(t)$, the infected $I(t)$, and the removed (recovered or deceased) $R(t)$ populations. It contains three parameters. Parameter $\beta$ is
the average number of contacts per person per time, multiplied by the probability of disease transmission in contact between a susceptible and an infectious subject, parameter $\gamma$ is the number of recovered or dead during one day divided by the total number of infected on that same day, and $N$ is the size of the population.

In the present COVID-19 outbreaks, the number of cases is a very unreliable variable, since it depends on the number of tests and the testing protocols, which are highly variable both in time and in
administrative districts. Similarly, the number of recovered patients
is dependent on the case definition, which in turn depends on the
testing protocol and frequency. The number of fatalities seems to be a better characteristic. Even if there can be differences in various administrative districts, how the deaths are counted, we can hope that
it is done consistently, and the relative numbers are accurate even if the total number can be debated. There is also some variability
in this data, and sometimes deaths appear in the statistics with a few days delay. Therefore, we would like to use the aggregated data to reduce the variation coming from this. 

In the next section, we show on the examples of the SIR and SEIR epidemic models\cite{Brauer} that they can be reduced to the dynamics of a single variable, the cumulative number of deaths.

\section{The dynamics of cumulative deaths in the SIR model}

The SIR model consists of three differential equations. The first one
expresses the relation between the number of removed (recovered or dead)
and the number of infected as
\begin{equation}
    \frac{dR(t)}{dt}=\gamma I(t).\label{R}
\end{equation}
It is assumed that the number of recovered and the number of deaths have a fixed ratio
\begin{equation}
    D(t)=p R(t),\label{deaths}
\end{equation}
where $p=p_D/(1-p_D)$ and $p_D$ is the probability of death, and $D(t)$ is the cumulative number of fatalities from the beginning of the outbreak. We would like to formulate our epidemic equations in terms of this variable.

The second equation expresses the relationship between the susceptible and the infected populations
$$
    \frac{dS(t)}{dt}=-\beta \frac{S(t)}{N}\cdot I(t).
$$
Using (\ref{deaths}) and (\ref{R}) we can write this in terms of deaths
as 
$$
    \frac{1}{S(t)}\frac{dS(t)}{dt}=-\frac{\beta}{\gamma pN}\frac{dD(t)}{dt},
$$
and can be solved to yield
\begin{equation}
    S(t)=N\exp\left(-\frac{\beta D(t)}{\gamma p N}\right),\label{S}
\end{equation}
where we assumed that at the beginning the whole population was susceptible $S(0)=N$ and nobody died of yet $D(0)=0$.

\begin{figure}[htb]
\centering
\includegraphics[width=11cm]{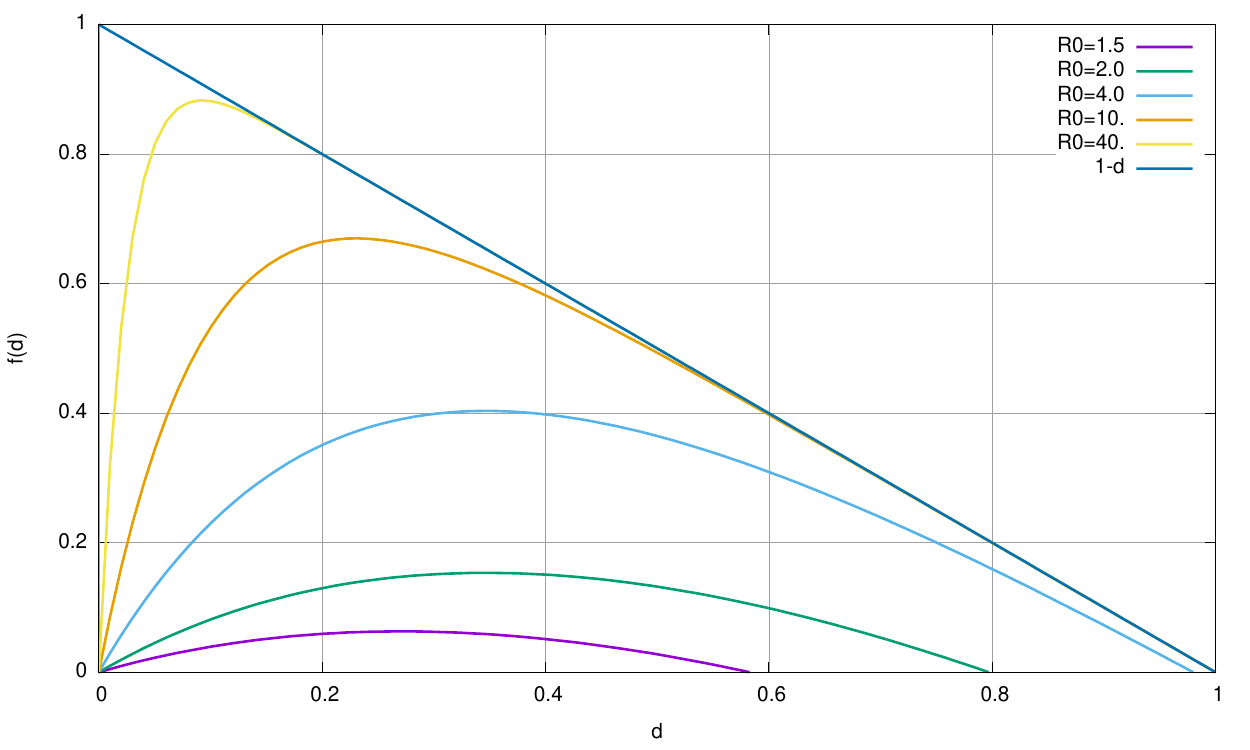}
\caption{The dimensionless function $f(d)=(1-e^{-R_0d})-d$ for various basic reproduction number parameters $R_0$. For large $R_0$ the function gets close to
its asymptotic form $f(d)=1-d$.
\label{fig1}}
\end{figure}

The third equation accounts for the time evolution of the infected part of the population
$$
    \frac{dI(t)}{dt}=\beta \frac{S(t)}{N}\cdot I(t)-\gamma I(t).
$$
We can use (\ref{S}),(\ref{R}) and (\ref{deaths}) and write this equation in terms of the cumulative number of fatalities 
$$
    \frac{d^2D(t)}{dt^2}=\left[\beta \exp\left(-\frac{\beta D(t)}{\gamma p N}\right) -\gamma \right]\frac{dD(t)}{dt}.
$$
The right hand side of this equation can be simplified to
$$
    \frac{d^2D(t)}{dt^2}=\frac{dF(D(t))}{dt},
$$
where we introduced
$F(D)=N\gamma p(1-e^{-\beta D/\gamma p N})-\gamma D$. Integrating both sides and setting the boundary condition $dD/dt=0$ for $D=0$ we get a
first order differential equation
$$
    \dot{d}(\tau)=f(R_0,d(\tau)),
$$
where we introduced the normalized deaths $d(t)=D(t)/N_{D}$, 
$N_{D}=pN$ is the number of deaths that would occur if the whole population would be infected and the dimensionless
time variable $\tau=\gamma t$. The parameter $R_0=\beta/\gamma$ is the basic reproduction number of the SIR model, and we introduced 
the dimensionless function
$f(R_0,d)=(1-e^{-R_0d})-d$, shown in Fig.\ref{fig1}. This equation has already been derived in Ref.\cite{sir}.

The most important aspect of this finding is that the cumulative number of deaths is a single, robust and reliable master variable, and the evolution of
the epidemic can be visualized in the two-dimensional phase space $(\dot{D}(t),D(t))$. Moreover, on a daily resolution of $\dot{D}(t)=D(t)-D(t-1)$ is
the daily increment, which is the daily number of fatalities.
In the next section we 

\section{Application of the model for the outbreaks in progress }

To verify the model, we looked at the available data\cite{JHU}. The function $f(d)$ is single-humped as one can see on Fig.~\ref{fig1}. To get data points
right from its peak, we have to look at examples where the outbreak passed its peak (at the time of the writing on April 17, 2020). We have found only three cases where this already happened with great certainty: China, Italy, and Spain.
In these countries, a lockdown has been implemented, and the disease can spread only in a small fraction of society, where infected people are locked down with
susceptible ones. The rest of society is not in contact with the epidemic. The number of people affected is $N$, and if all of them get infected, $N_{D}=pN$ people expected to die. The value of $p$ is unknown. It is estimated\cite{Nishiura} from the data of countries, where there is extensive testing in places
such as South Korea ($p\approx 0.02$) and Iceland ($p\approx 0.004$).
We expect to extract the parameters $\beta$, $\gamma$ and the number of people
in danger of dying $D_{max}$.

\begin{figure}[htb]
\centering
\includegraphics[width=11cm]{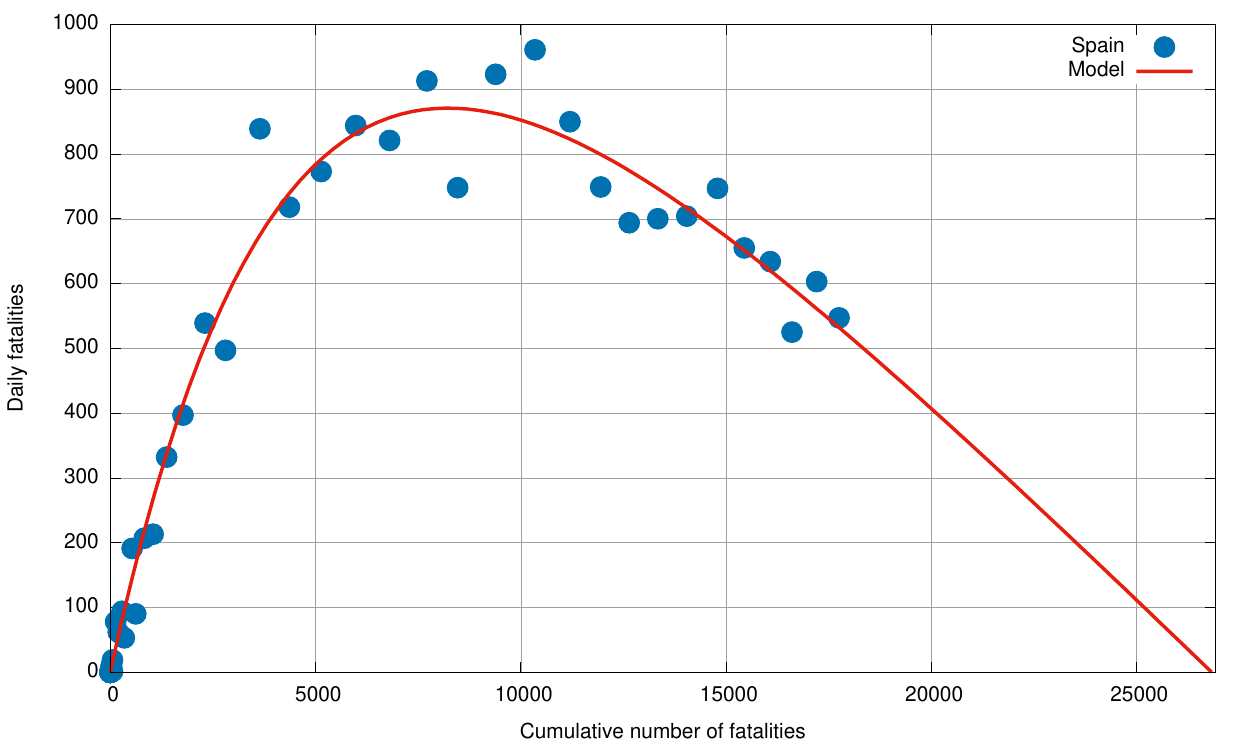}
\includegraphics[width=11cm]{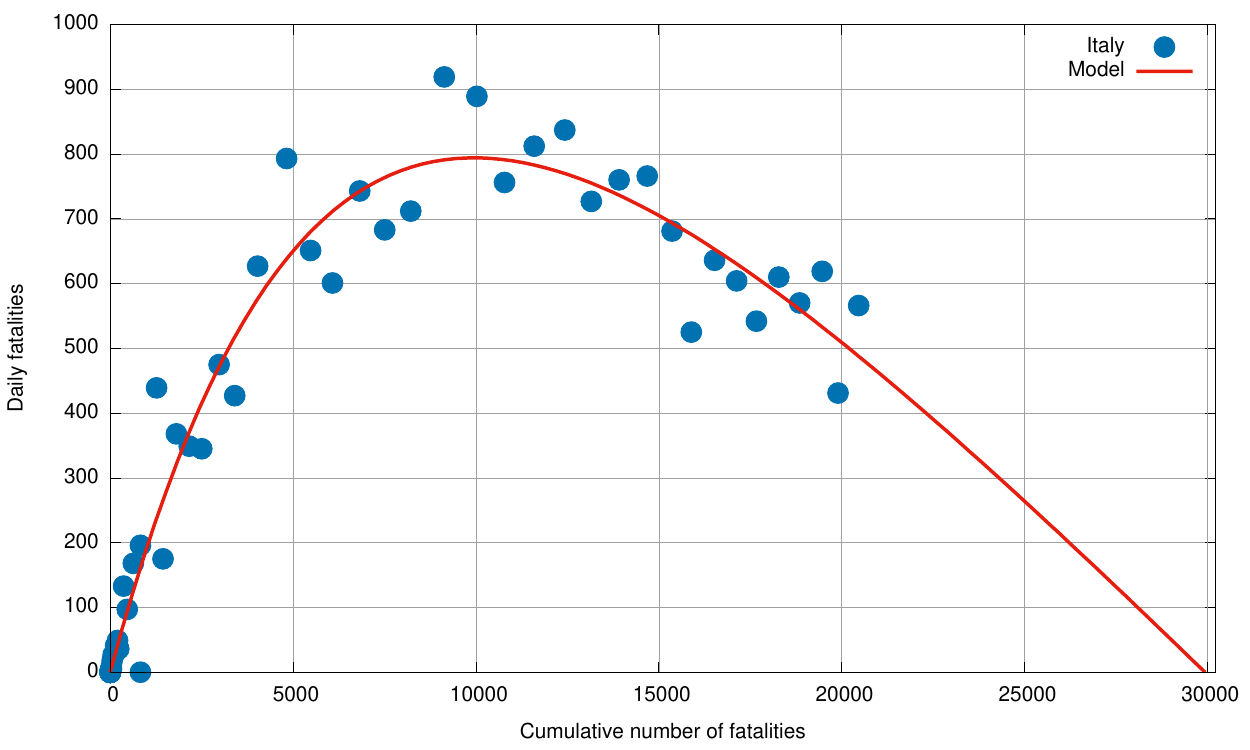}
\includegraphics[width=11cm]{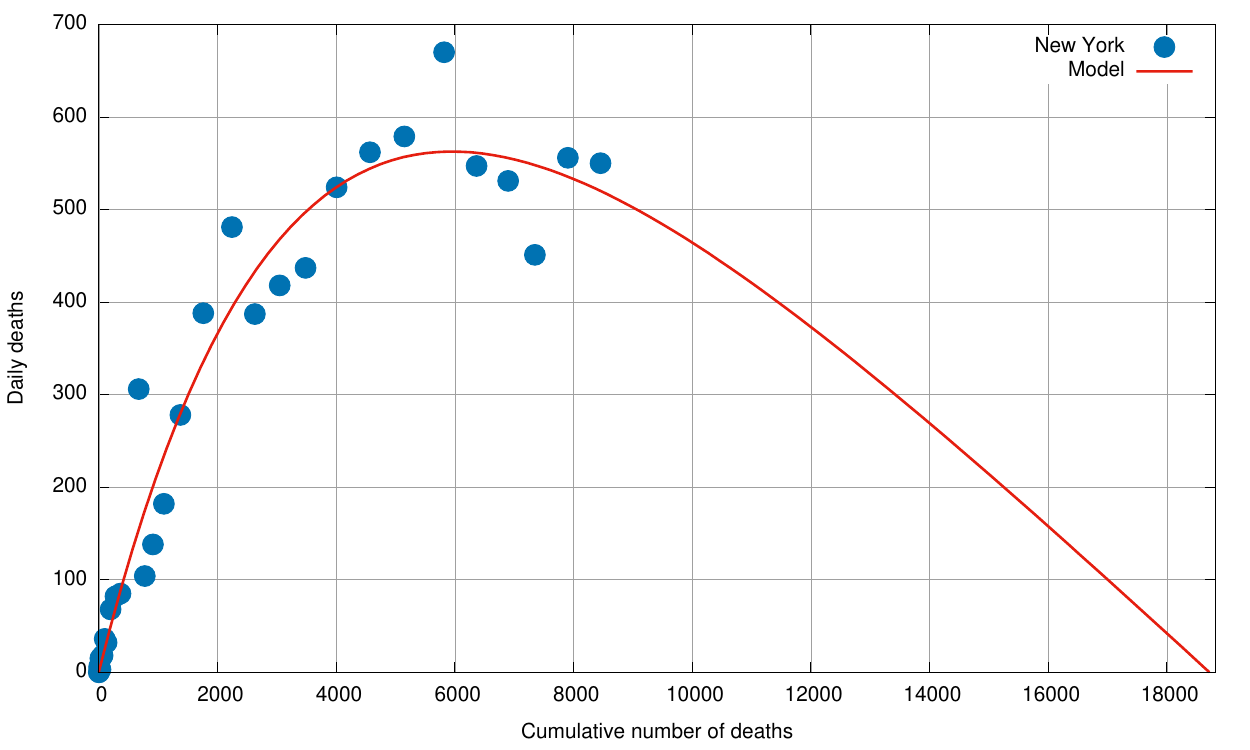}
\caption{Daily number of deaths as a function of the cumulative number of deaths for Spain, Italy, and New York. Data ends on April 13 for Spain and Italy and on April 15 for New York. The red curve is $K(1-e^{-\beta D/K})-\gamma D$ with parameters in Table 1.
\label{fig2}}
\end{figure}

In Fig.~\ref{fig2}, we show the daily number of fatalities as a function of cumulative deaths for Spain, Italy, and New York. We also fit our model to the data and show the parameters in Table~\ref{tab:pars}.
The parameters are very similar for all three cases. The parameter $1/\gamma\approx 17\; days$ is in good agreement with the typical length of the disease till recovery or death. The basic reproduction number $R_0\approx 5$ is
somewhat elevated, but it is also realistic for people in close contact with the infected (for example, family members and doctors) in the lockdown situation.
According to the model, the lockdown works as expected.
We have to stress, that in absence of a complete lockdown and social distancing, this outcome is not guaranteed and the epidemic can change its course when social distancing is lifted.
\begin{table}[htb]
    \centering
    \begin{tabular}{c|c|c|c|c}
       Parameter  &  Spain & Italy& New York &Hubei\\ \br
        $K=\gamma p N$ & $1677 \pm 151$& $1733  \pm 285 $& $1141\pm 703$ &$694 \pm 210$\\
        $\beta$ & $0.355 \pm 0.008$ & $0.269 \pm 0.008$& $0.32\pm 0.04$ &$0.369 \pm 0.029$\\
        $\gamma$ & $0.062 \pm 0.008$ & $0.0573 \pm 0.012$& $0.060\pm 0.06$ & $0.177 \pm 0.038 $\\
        $R_0=\beta/\gamma$& $\approx 5.7$ & $\approx 4.7$& $\approx 5.3$ &$\approx 2.1$\\
        $N_{D}=K/\gamma$& $\approx 27,000$& $\approx 30,000$& $\approx 19,000$&$\approx 3900$\\ 
    \end{tabular}
    \caption{Parameters of fitting $K(1-e^{-\beta D/K})-\gamma D$ for Spain,  Italy, New York and Hubei Province, China. The basic reproduction number $R_0=\beta/\gamma$ and the number of
    people in danger of dying $N_D$ is calculated from the most likely values.  }
    \label{tab:pars}
\end{table}

\begin{figure}[htb]
\centering
\includegraphics[width=11cm]{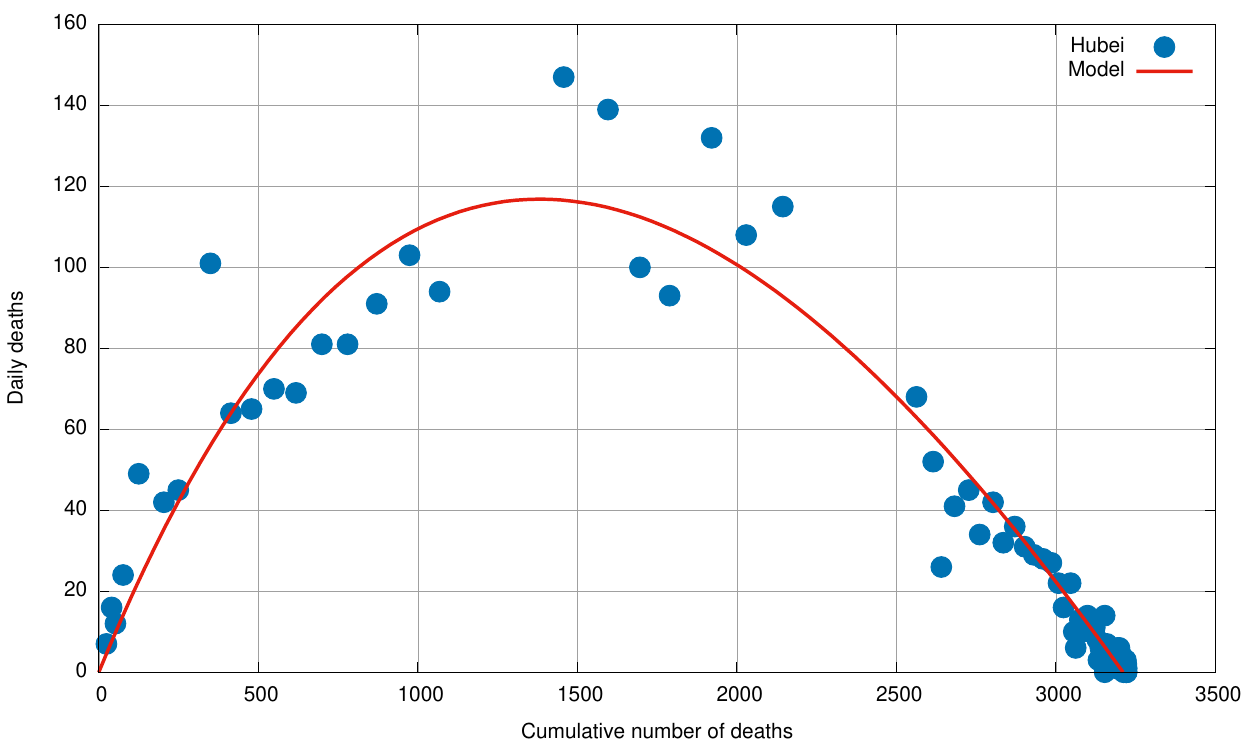}
\caption{Daily number of deaths as a function of the cumulative number of deaths for Hubei Province, China. Data ends on April 13, 2020. The red curve is $K(1-e^{-\beta D/K})-\gamma D$ with parameters in Table 1.
\label{fig3}}
\end{figure}

Next, we look at the data from Hubei Province, China. Due to irregularities in the reporting, we removed data points where zero case has been reported, and then the missing cases two were reported the next day. In Fig.~\ref{fig3} we show the data and the fitted parameters are in Table\ref{tab:pars}. The appearance of the model curve and the parameters are very different from those observed in the European countries. 
The theoretical curve is an almost symmetric parabola, which is the mathematical consequence of the low basic reproduction number $R_0\approx 2.1$, as we show later. The parameter
$\gamma$ is also substantially different from the values observed in Europe. Accordingly, the expected number of fatalities is smaller there as well. 

What can be the explanation for these differences in the parameters? For example, the recovery time $1/\gamma\approx 6\; days$ is much shorter than the natural recovery period observed in Europe and New York. The likely explanation is that the lockdown in China has been implemented differently. Once someone tested positive or symptoms appeared, they got immediately separated and isolated in designated quarantine facilities. This way, the period in which the patient could infect others has been reduced to an average of 6 days from 17 days in Europe and New York. The basic reproduction number fell accordingly. 

\section{Forecasting the outcome}

It is vital to gain information about the course of the epidemic during the outbreak to monitor the effectiveness of policy actions. The first question is the total number of fatalities $D_T$. In the present model, it is given by the non-trivial zero $F(D_\infty)=0$. The solution depends only on the basic reproduction number and on $N_{D}=pN$
and given by the solution of
\begin{equation}
    1-D_\infty/N_{D}=e^{-R_0D_\infty/N_{D}}.
\end{equation}
This equation can be solved numerically. There are two important limiting cases where we can solve the equations explicitly, and we can use these solutions approximately. One is when $R_0$ is small, and the other is when it is large. 
We can use these as an approximate solution when $1<R_0<2.3$ and when $R_0>2.3$, respectively.
We analyze these next.

\subsection{Logistic growth}

When $0<x<0.7$ the exponential function can be approximated within 10\% error with the quadratic terms in the expansion $\exp(-x)\approx 1-x+x^2/2$. Using this the differential
equation becomes
\begin{equation}
    \frac{dD}{dt}=(\beta-\gamma) D \left[1-\frac{R_0^2}{2(R_0-1)N_D}D\right].
\end{equation}
The right hand side is a parabola and this is the logistic growth equation. 
The final number of fatalities is $D_\infty = N_D2(R_0-1)/R_0^2$.
This approximation describes the situation in Hubei well. The parameters of the logistic
equation can be easily determined by fitting a linear regression line to the daily number of relative growth $(D(t)-D(t-1))/D(t)$ as a function of $D(t)$. 
\begin{figure}[htb]
\centering
\includegraphics[width=11cm]{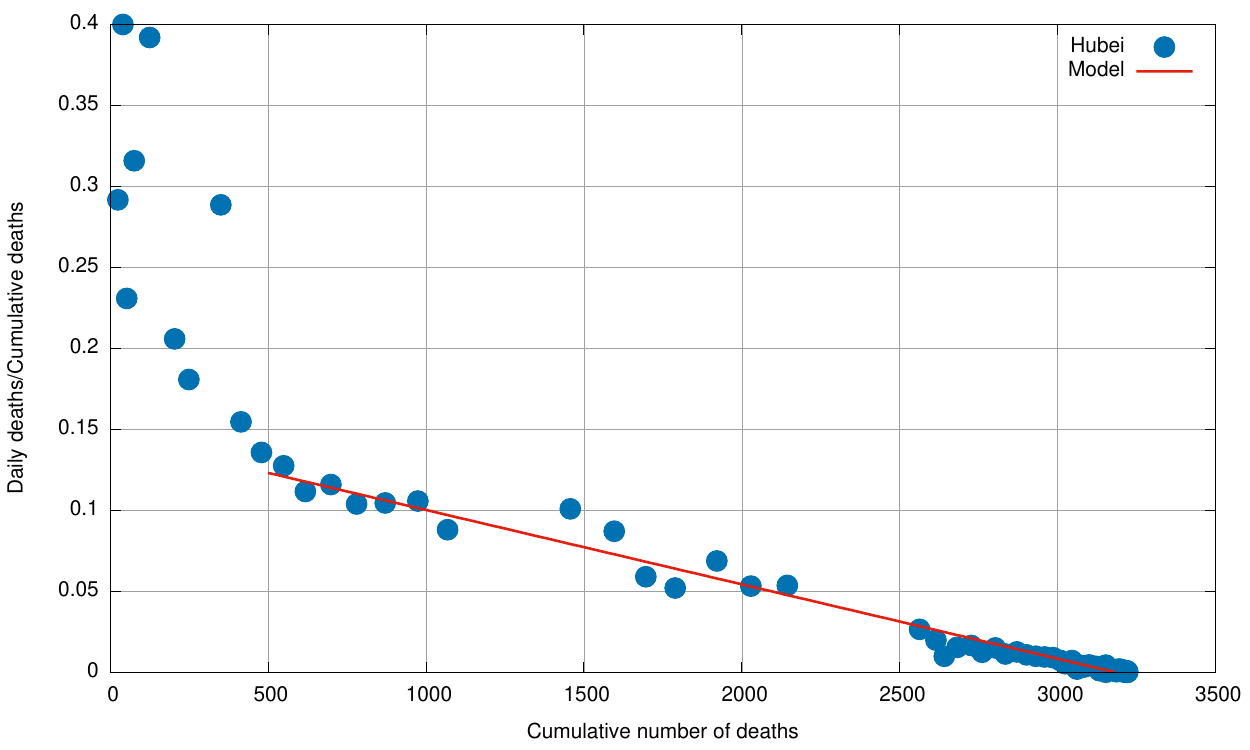}
\caption{Relative growth rate of fatalities (Daily deaths/Cumulative number of deaths) for the outbreak in Hubei Province, China. The red regression line is fitted from the cumulative death of 500.
\label{fig4}}
\end{figure}
In Fig.~\ref{fig4}. we show the relative growth vs. cumulative death for Hubei. After the initial transients, one can predict the final number of fatalities by continuing the linear trend in the data. 

In the general case the maximum number of daily deaths occurs when $dF(D^*)/dD=0$ . This equation can be solved in the general case and yields $D^*=N_D\log(R_0)/R_0$. In the small $R_0$ case $\log(R_0)=-\log(1/R_0)=-\log(1+1/R_0-1)\approx 1-1/R_0$ and $D^*=N_D(R_0-1)/R_0^2$. In the logistic growth regime the midpoint of the epidemic is at the height of the epidemic $D^*=D_\infty/2 $.

\subsection{The exponential regime}

\begin{figure}[htb]
\centering
\includegraphics[width=11cm]{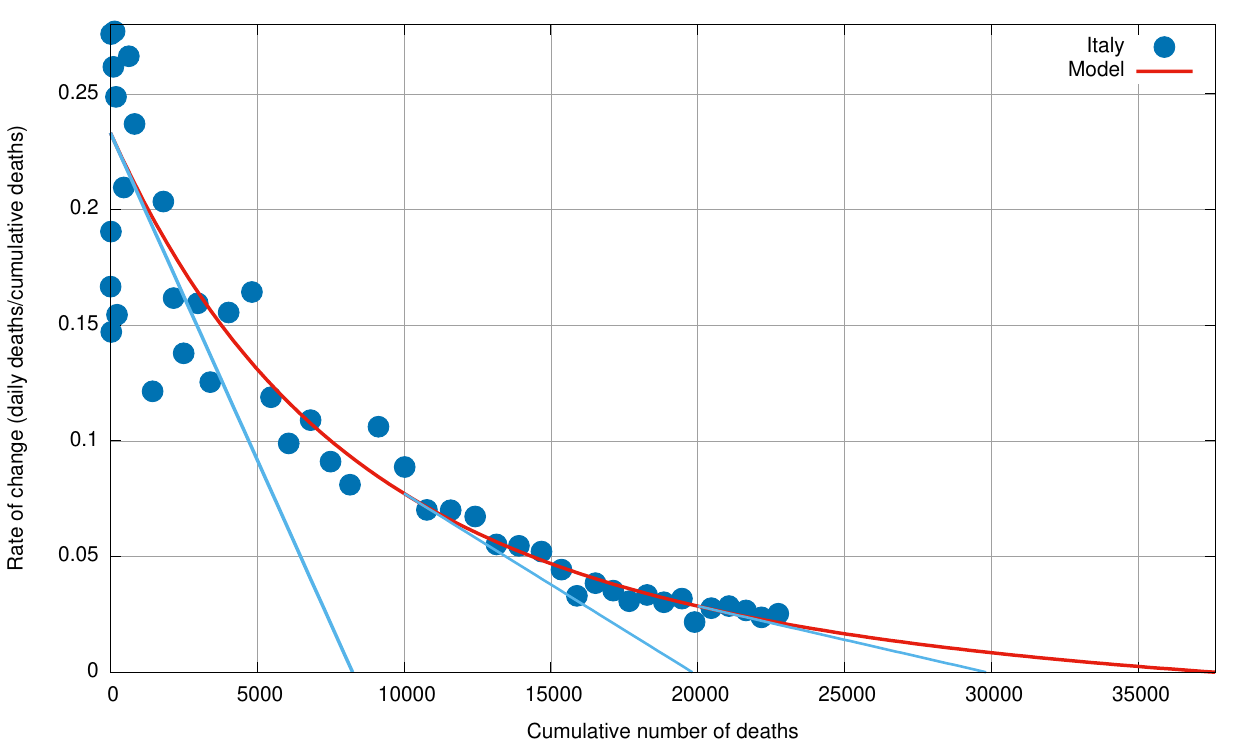}
\caption{Relative growth rate of fatalities (Daily deaths/Cumulative number of deaths) for the outbreak in Italy. The red line is the fitted SIR model with parameters in Table 1. The blue lines represent the linear trend continuation from 0, 10,000, and 20,000 fatalities.
\label{fig5}}
\end{figure}

\begin{figure}[htb]
\centering
\includegraphics[width=11cm]{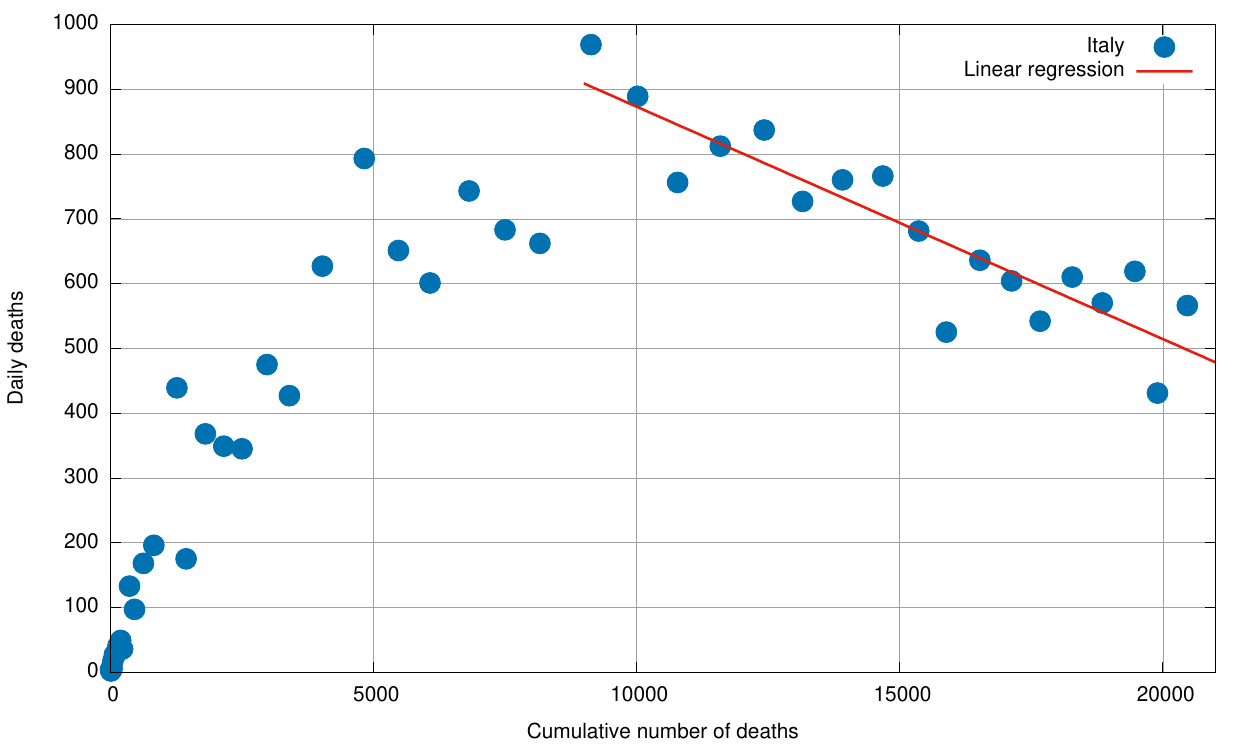}
\caption{Daily number of fatalities for the outbreak in Italy. The red line is the fitted linear regression $\gamma(N_D-D)$ with parameters $\gamma=0.036 \pm 0.004$ and $N_D=34335 \pm 2331$.
\label{fig6}}
\end{figure}

When the basic reproduction number is large, the logistic grow model breaks down. In Fig.~\ref{fig5} we demonstrate this on the data from Italy, where $R_0\approx 4.7$. 
While the SIR model describes
the evolution correctly, linear trend continuation leads to larger and larger estimates for the number of fatalities.
A better approximation can be developed by assuming that after the peak of the epidemic ($D^*\ll D$), the argument of the exponential is large, and the term can be dropped. This 
approximation leads to 
\begin{equation}
    \frac{dD}{dt}=\gamma (N_D-D).
\end{equation}
We can fit this linear model to the data from Italy after the peak of the epidemic.
The result is in Fig.~\ref{fig6}. The linear regression is in good agreement with the result of the full SIR model.
This approximation describes the situation when in the SIR model, the susceptible population
is small, and the number of infected decreases exponentially with the rate of $\gamma$. 
The recovered and deceased population saturates exponentially at the same rate.
Since the recovery time $T_r=1/\gamma$ is large, this recovery process is much slower than the buildup of the
epidemic, which is happening on the timescale of $T=1/(\beta-\gamma)=T_r/(R_0-1)$.
In the Italian case, the recovery period after the peak is about four times longer than
the time from the beginning to the peak.

\section{Summary and outlook}

In the previous sections, we have seen that the simple one-dimensional differential equation derived from the SIR model describes the data quite reliably. The SIR model is not able to capture all the details of the epidemic process, so the question is why it is a good approximation in the present situation. To see this, we take a more complicated and more adequate model to see why it can be reduced to an effective SIR model. 

\begin{figure}[htb]
\centering
\includegraphics[width=7cm]{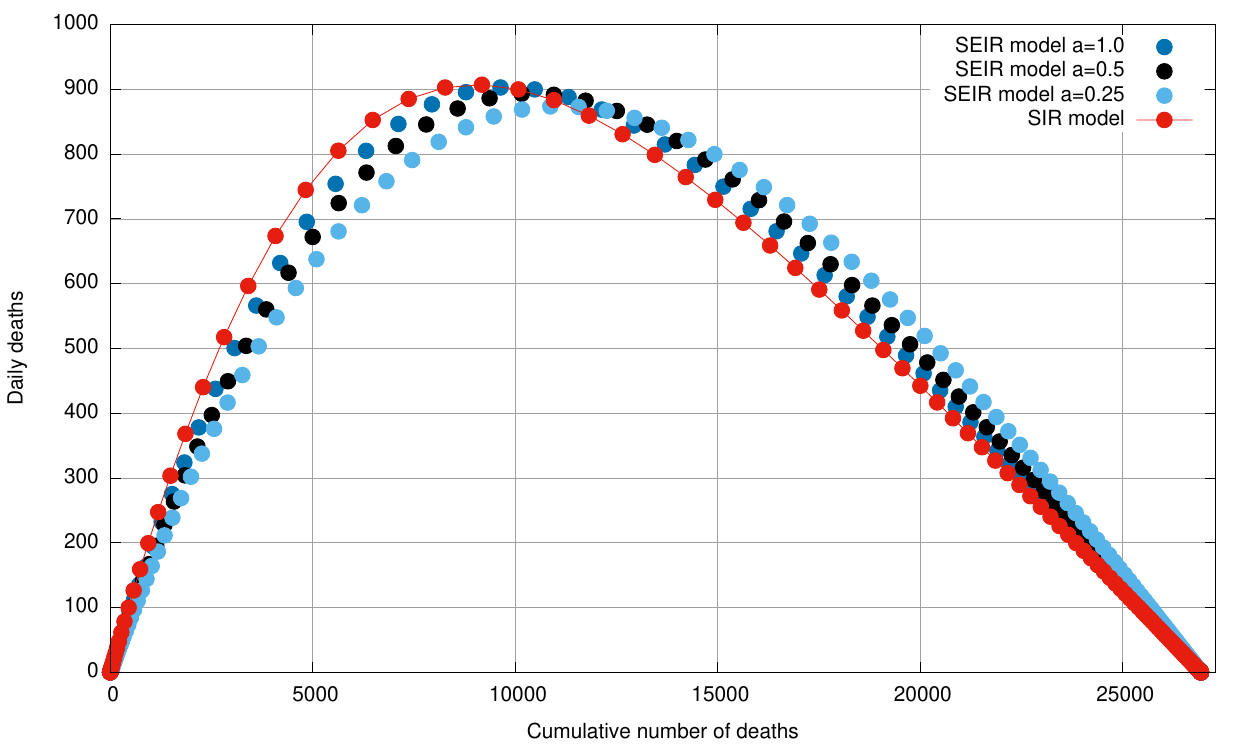}
\includegraphics[width=7cm]{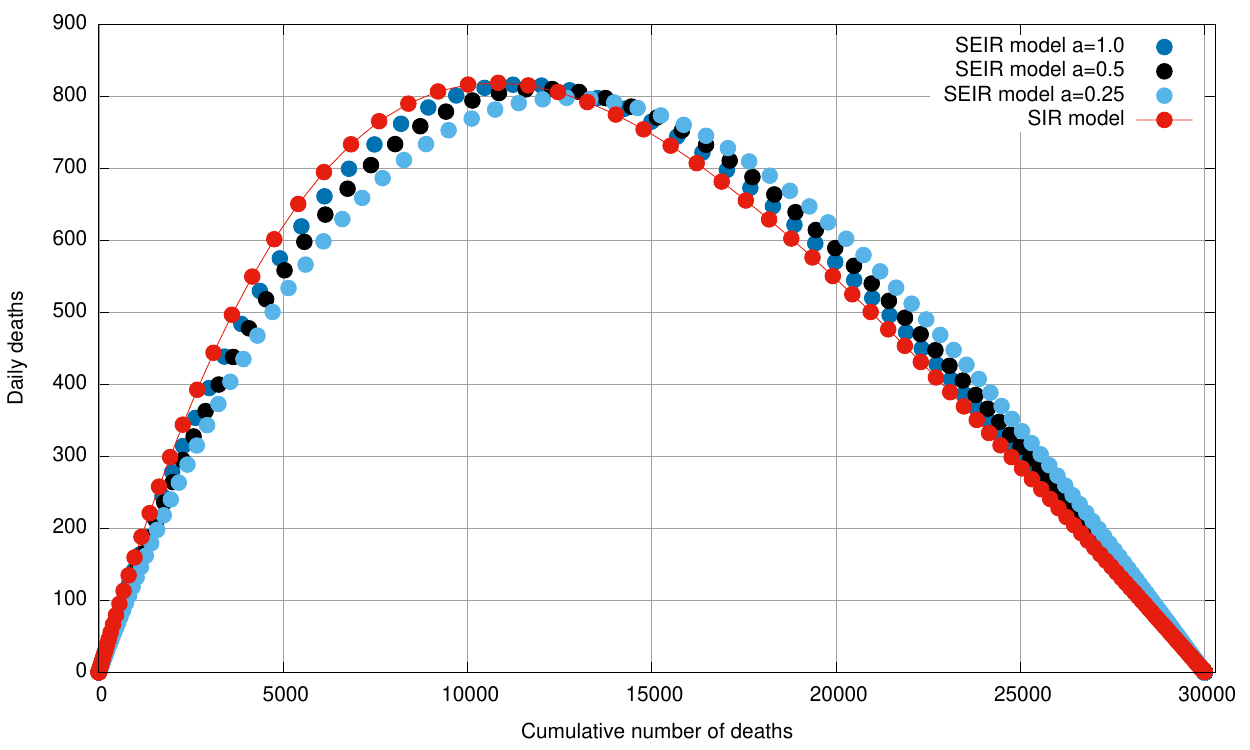}
\includegraphics[width=7cm]{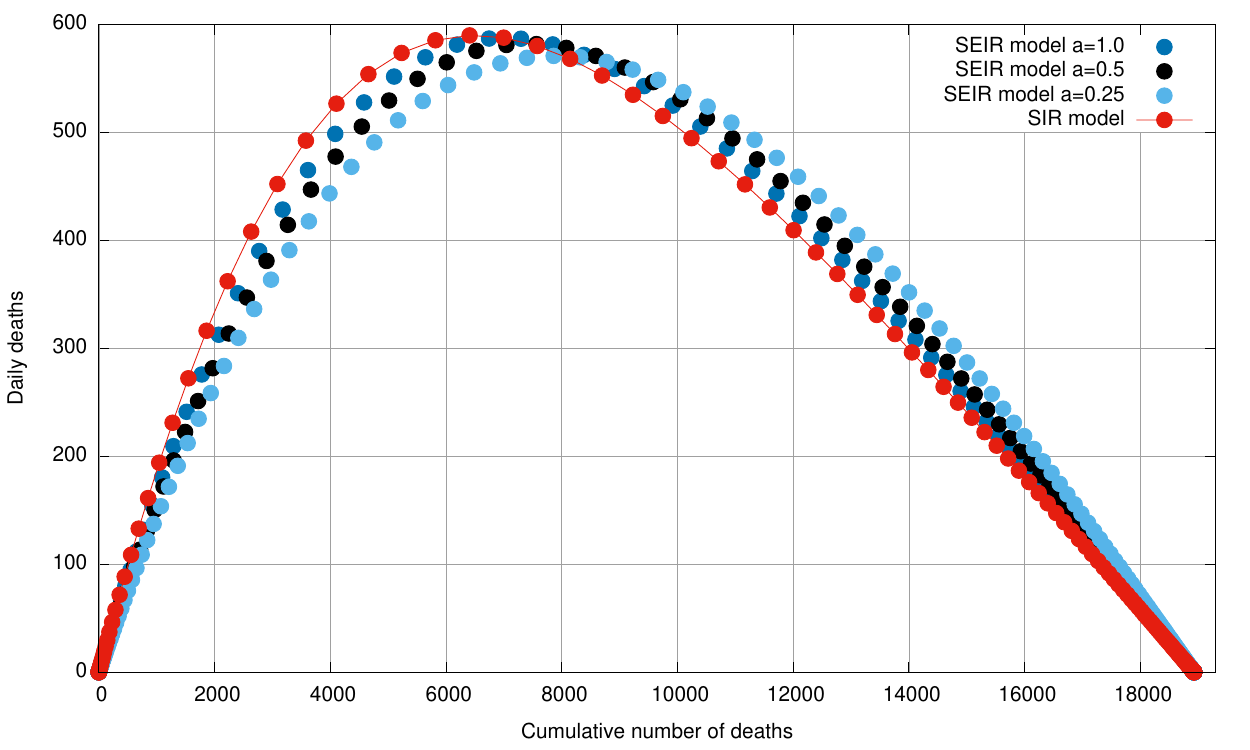}
\includegraphics[width=7cm]{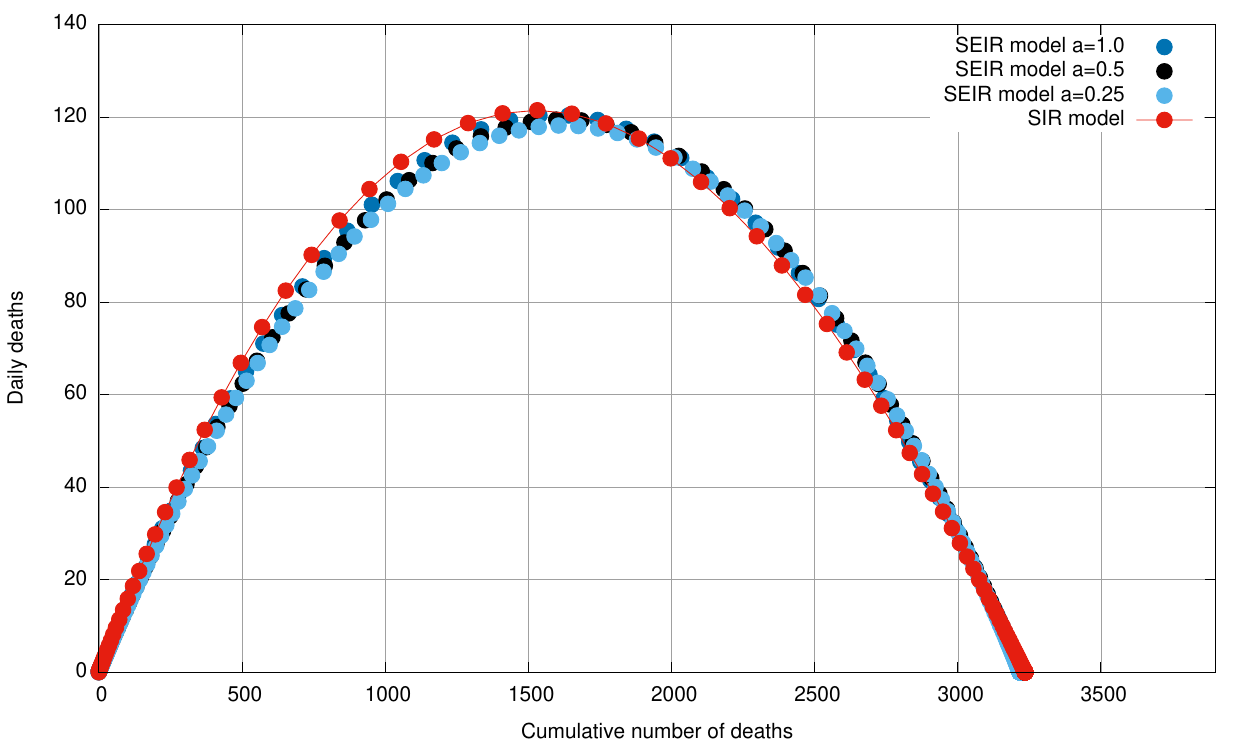}
\caption{Daily number of deaths as a function of the cumulative number of deaths generated from the parameters of Table 1. for Spain, Italy, New York and Hubei using the SIR model (red dots) and the SEIR model with incubation period of one day $a=1$, two days $a=0.5$ and four days $a=0.25$. In the SEIR model the number of deaths and the parameter $N_D$ are scaled with the factor $1/(1+\gamma/a)$. The starting and ending slopes of the curves match exactly due to the mathematical relation between the models, while the tops of the curves show slight dependence on $a$.
\label{fig7}}
\end{figure}

The SEIR model takes into consideration the incubation period, when the person is already infected but still not infecting others. The equation for the number of 
exposed is
$$
\frac{dE(t)}{dt}=\beta\frac{S(t)}{N}I(t)-aE(t),\label{exposed}
$$
and for the number of infectious is
$$\frac{dI(t)}{dt}=aE(t)-\gamma I(t), $$
where $T_I=1/a$ is the average incubation period. 
From this last equation, we can express the number of exposed with the number of infectious
$aE(t)=dI(t)/dt+\gamma I(t)$ and using this expression, we can rewrite (\ref{exposed}) the following way
$$\frac{1}{a}\frac{d^2I(t)}{dt^2}+ \left[1+\gamma/a\right]\frac{dI(t)}{dt}=\beta\frac{S(t)}{N}I(t)-\gamma I(t).$$
If the parameter $a$ is larger than the two characteristic rates of the SIR model, $a\gg \beta$ and $a\gg \gamma$, then
the term with the second derivative of the number of infected is much smaller that the first derivative term and can be neglected $(1/a)d^2I(t)/dt^2\ll dI/dt$ in the equation and we recover the SIR equation
$$
    \frac{dI(t)}{dt}=\beta_e \frac{S(t)}{N}I(t)-\gamma_e I(t),
$$
with the new effective parameters $\beta_e=\beta/(1+\gamma/a)$ and $\gamma_e=\gamma/(1+\gamma/a)$.
The effective parameter $\gamma_e$ reflects the fact that the average time from infection to recovery or death changes from $1/\gamma$ to $1/\gamma_e=1/\gamma + 1/a$ due to the addition of the incubation time. The ratio $R_0=\beta/\gamma=\beta_e/\gamma_e$ remains unchanged. The solution becomes now $F(D)=N\gamma_ep(1-e^{-\beta_eD/\gamma_epN})-\gamma_eD=N'p\gamma (1-e^{-\beta D'/\gamma pN'})-\gamma D'$, where we
introduced the rescaled quantities $N'=N/(1+\gamma/a)$ and $D'=D/(1+\gamma/a)$. In other words, for sufficiently large parameter values $a$, a SEIR model simulation with parameters $\beta$, $\gamma$ and $N$ should collapse on the SIR simulation with the same $\beta$ and $\gamma$ parameters and $N'=N/(1+\gamma/a)$ once we re-scale the number of deaths according to $D'=D/(1+\gamma/a)$.
In Fig.~\ref{fig7}. we show the effect of the incubation period on the shape of the epidemic curve. Using the parameters of Table 1. we generate epidemic curves for Spain, Italy, New York and Hubei with the SEIR model for incubation periods $T_I=1/a=$ 1, 2 and 4 days. After re-scaling $N$ and $D$ the SEIR simulations nearly collapse on the SIR curve as expected from the previous discussion.  

Similarly, more complicated epidemic models can be reduced to an effective SIR model when the time scales of the variables are much shorter than those in the SIR model. In the case of the separation of timescales, the fast-changing variables can be eliminated via averaging, and a few slow master variables drive the process.

In this paper, we argued that the cumulative number of fatalities is the single master variable in the epidemic process, and its evolution can be described by a 
first-order differential equation. The model can be verified on the available data from the advanced outbreaks in Spain, Italy, New York, and Hubei Province, China.
The effective $\beta$ parameter of the process is very similar in all four cases,
$\bar{\beta}=0.33$ in average, The infectious period is about $1/\bar{\gamma}=16.7\; days$ is in Spain, Italy, and New York, while it is only 6 days in China, probably due to the policy of separation of the infected in designated quarantine areas.     

\ack

We thank fruitful discussions with István Csabai, Péter Pollner, Géza Meszéna and Viktor Müller. 
The research was financed by the Research Excellence Programme of the Ministry for Innovation 
and Technology in Hungary, within the framework of the Digital Biomarker thematic programme 
of the Semmelweis University and the ELTE Excellence Program (783-3/2018/FEKUTSRAT).

\section*{References}


\end{document}